\documentclass[10pt,a4paper]{article}

\usepackage{graphicx}
\usepackage{amsmath}
\usepackage{amsfonts}
\usepackage[capitalize,noabbrev]{cleveref}
\usepackage{amssymb}
\usepackage[T1]{fontenc}
\usepackage[utf8]{inputenc}
\usepackage{cite}
\usepackage{subfig}
\usepackage{amsmath}
\usepackage{algorithm2e}
\usepackage{todonotes}
\RequirePackage{acronym}
\acrodef{AODV}{Ad Hoc on Demand Distance Vector}
\acrodef{API}{Application Programming Interface}
\acrodef{ASC}{Access Service Class}
\acrodef{ASS}{Anonymity Set Size}
\acrodef{ATB}{Adaptive Traffic Beacon}
\acrodef{AHS}{Automated Highway Systems}
\acrodef{AV}{Autonomous Vehicle}
\acrodef{AWGN}{Additive White Gaussian Noise}
\acrodef{BSM}{Basic Safety Message}
\acrodef{C2C}{Car-to-Car}
\acrodef{C2I}{Car-to-Infrastructure}
\acrodef{CAM}{Cooperative Awareness Message}
\acrodef{CANU}{Communication in Ad Hoc Networks for Ubiquitous Computing}
\acrodef{CA}{Certificate Authority}
\acrodef{CCA}{Clear Channel Assessment}
\acrodef{CCH}{Control Channel}
\acrodef{DSP}{Digital Signal Processor}
\acrodef{CDF}{Cumulative Distribution Function}
\acrodef{CPU}{Central Processing Unit}
\acrodef{COTS}{Commercial Off-The-Shelf}
\acrodef{CPRNG}{Cryptographic Pseudorandom Number Generator}
\acrodef{CRL}{Certificate Revocation List}
\acrodef{CoCar}{Cooperative Cars}
\acrodef{CV}{Conventional Vehicle}
\acrodef{DCC}{Decentralized Congestion Control}
\acrodef{DCH}{Dedicated Channel}
\acrodef{DENM}{Decentralized Environmental Notification Message}
\acrodef{DHT}{Distributed Hash Table}
\acrodef{DSC}{DCC Sensitivity Control}
\acrodef{DSRC/WAVE}{Dedicated Short-Range Communication~/ Wireless Access in the Vehicular Environment}
\acrodef{DSRC}{Dedicated Short-Range Communication}
\acrodef{DSR}{Dynamic Source Routing}
\acrodef{DTN}{Delay Tolerant Network}
\acrodef{DYMO}{Dynamic MANET On Demand}
\acrodef{ECDF}{Empirical Cumulative Distribution Function}
\acrodef{ECU}{Electronic Control Unit}
\acrodef{EDCAF}{Enhanced Distributed Channel Access Function}
\acrodef{EDCA}{Enhanced Distributed Channel Access}
\acrodef{FACH}{Forward Access Channel}
\acrodef{FOT}{Field Operational Test}
\acrodef{FIFO}{First In, First Out}
\acrodef{GGSN}{Gateway GPRS Support Node}
\acrodef{GLOSA}{Green Light Optimal Speed Advisory}
\acrodef{GPS}{Global Positioning System}
\acrodef{ICWS}{Intersection Collision Warning System}
\acrodef{IDM-IM}[IDM\_IM]{IDM with Intersection Management}
\acrodef{IDM-LC}[IDM\_LC]{IDM with Lane Changing}
\acrodef{IDM/MOBIL}{Intelligent-Driver/MOBIL Model}
\acrodef{IDM}{Intelligent Driver Model}
\acrodef{ITS}{Intelligent Transportation System}
\acrodef{IVC}{Inter-Vehicle Com\-mu\-ni\-ca\-tion}
\acrodef{JPDA}{Joint Probabilistic Data Association}
\acrodef{LBU}{Last Before Unavoidable}
\acrodef{LLC}{Logical Link Control}
\acrodef{LOS}{Line of Sight}
\acrodef{LTE}{Long-Term Evolution}
\acrodef{MAC}{Medium Access Control Layer}
\acrodef{MANET}{Mobile Ad-Hoc Network}
\acrodef{MBMS}{Multimedia Broadcast/Multicast Service}
\acrodef{NLOS}{Non-Line-of-Sight}
\acrodef{MMTS}{multi-agent microscopic traffic simulator}
\acrodef{MOBIL}{Minimizing Overall Braking Induced by Lane change}
\acrodef{MOVE}{Mobility model generator for Vehicular networks}
\acrodef{NIC}{Network Interface Controller}
\acrodef{OBU}{On-Board Unit}
\acrodef{OFDM}{Orthogonal Frequency Division Multiplexing}
\acrodef{PATH}{Partners for Advanced Transit and Highways}
\acrodef{PAPR}{Peak to Average Power Ratio}
\acrodef{PRNG}{Pseudorandom Number Generator}
\acrodef{PDN}{Public Data Network}
\acrodef{PDU}{Protocol Data Unit}
\acrodef{PET}{Privacy-Enhancing Technology}
\acrodef{PHY}{Physical Layer}
\acrodef{PKI}{Public Key Infrastructure}
\acrodef{PTM}[p-t-m]{Point-to-Multipoint}
\acrodef{PTP}[p-t-p]{Point-to-Point}
\acrodef{QPSK}{Quadrature Phase-shift Keying}
\acrodef{QoS}{Quality of Service}
\acrodef{RACH}{Random Access Channel}
\acrodef{RAN}{Radio Access Network}
\acrodef{ROI}{Region of Interest}
\acrodef{RSU}{Roadside Unit}
\acrodef{SCH}{Service Channel}
\acrodef{SDR}{Software Defined Radio}
\acrodef{SIFS}{Short Interframe Spacing}
\acrodef{SINR}{Signal-to-Interference-plus-Noise Ratio}
\acrodef{SNR}{Signal to Noise Ratio}
\acrodef{SOTIS}{Self-Organizing Traffic Information System}
\acrodef{SPAT}{Signal Phase and Timing Message}
\acrodef{SSU}{Stationary Support Unit}
\acrodef{STRAW}{Street Random Waypoint}
\acrodef{SUMO}{Simulation of Urban Mobility}
\acrodef{TAC}{Transmit Access Control}
\acrodef{TDC}{Transmit Data Rate Control}
\acrodef{TIC}{Traffic Information Center}
\acrodef{TIS}{Traffic Information System}
\acrodef{TOPO}{Road Topology Message}
\acrodef{TPC}{Transmit Power Control}
\acrodef{TPEG}{Transport Protocol Expert Group}
\acrodef{TRC}{Transmit Rate Control}
\acrodef{TSTM}{Traffic Signal Timing Manual}
\acrodef{TTL}{Time-to-Live}
\acrodef{TraCI}{Traffic Control Interface}
\acrodef{UE}{User Equilibrium}
\acrodef{UML}{Unified Modeling Language}
\acrodef{UMTS}{Universal Mobile Telecommunications System}
\acrodef{V2I}{Vehicle-to-Infrastructure}
\acrodef{V2V}{Vehicle-to-Vehicle}
\acrodef{VANET}{Vehicular Ad-Hoc Network}
\acrodef{VTL}{Virtual Traffic Light}
\acrodef{WAVE}{Wireless Access in Vehicular Environments}
\acrodef{WSMP}{Wave Short Message Protocol}

\usepackage{authblk}
\usepackage[margin=2cm]{geometry}

\begin{document}

\title{Potentials and Implications of Dedicated Highway Lanes for Autonomous Vehicles}

\author[1]{Jordan Ivanchev}
\author[2]{Alois Knoll}
\author[1]{Daniel Zehe}
\author[1]{Suraj Nair}
\author[1]{David Eckhoff}
\affil[1]{TUMCREATE, Singapore}
\affil[2]{Department of Informatics, Technische Universit\"{a}t M\"{u}nchen, Germany}

\maketitle

\begin{abstract}

	The introduction of \acp{AV} will have far-reaching effects on road traffic in cities and on highways.The implementation of \ac{AHS}, possibly with a dedicated lane only for \acp{AV}, is believed to be a requirement to maximise the benefit from the advantages of \acp{AV}.
	We study the ramifications of an increasing percentage of AVs on the traffic system with and without the introduction of a dedicated AV lane on highways.
	We conduct an analytical evaluation of a simplified scenario and a macroscopic simulation of the city of Singapore under user equilibrium conditions with a realistic traffic demand. 
	We present findings regarding average travel time, fuel consumption, throughput and road usage.
	Instead of only considering the highways, we also focus on the effects on the remaining road network.
	Our results show a reduction of average travel time and fuel consumption as a result of increasing the portion of AVs in the system.
	We show that the introduction of an AV lane is not beneficial in terms of average commute time.
	Examining the effects of the AV population only, however, the AV lane provides a considerable reduction of travel time ($\approx 25\%$) at the price of delaying conventional vehicles ($\approx 7\%$).
	Furthermore a notable shift of travel demand away from the highways towards major and small roads is noticed in early stages of AV penetration of the system.
	Finally, our findings show that after a certain threshold percentage of AVs the differences between AV and no AV lane scenarios become negligible.

\end{abstract}
	
\section{Introduction}
Autonomous vehicles have ceased to be only a vision but are rapidly becoming a reality as cities around the world such as Pittsburgh, San Francisco, and Singapore have begun investigating and testing autonomous mobility concepts~\cite{henderson2016autonomous}.
The planned introduction of thousands of autonomous taxis, as currently planned in Singapore~\cite{henderson2016autonomous}, poses a challenge not only to the in-car systems of the \ac{AV}, but also to the traffic system itself.

Besides the deployment of more efficient ride sharing systems and the reduction of the total number of vehicles on the road, \acp{AV} can  traverse a road faster while using less space.
To achieve the maximum benefit in terms of traffic speeds and congestion reduction, however, mixing of \acp{AV} and \acp{CV} should be avoided~\cite{litman2016autonomous}.
One method to achieve this is the introduction of dedicated \ac{AV} lanes on highways to allow \acp {AV} to operate more efficiently due to the absence of unpredictable random behaviour introduced by humans and the use of communication capabilities, e.g. platoon organization~\cite{segata2014supporting}.

Blocking certain lanes for \acp{CV}, and thereby limiting the overall road capacity for human drivers is certainly a step that can have considerable ramifications, depending on the portion of \acp{AV} in the system.
While at the early stages, where only few \acp{AV} are on the road, it would constitute an incentive to obtain an \ac{AV}, it could also possibly generate traffic congestion and increase travel times for other vehicles ~\cite{ivanchev2017critical}.
As the level of \ac{AV} penetration in the road transportation system is increased, the total congestion level would likely drop, however, the advantage of using \acp{AV} over \acp{CV} would gradually be diminished as well. 

Lastly, converting highways into \ac{AHS} could also affect the rest of the road network, as drivers of \acp{CV} may then choose a different route, caused by the changed capacity of the highway, which can lead to a mismatch between road network and traffic demand ~\cite{7313331}.

In this paper, we take a closer look at the benefits and drawbacks of introducing a dedicated \ac{AV} lane on major highways Singapore.
Focusing particularly on the effect of varying percentages of \acp{AV} in the system, we study the differences in terms of capacity, travel time, fuel consumption, and impact on other roads compared to a setting where no dedicated \ac{AV} lane is assigned.
In short, our two main contributions are:
\begin{itemize}
\item We present an analytical evaluation of the expected benefit from the introduction of dedicated \ac{AV} lanes.
\item Using a macroscopic traffic simulation of the city-state of Singapore and based on realistic travel demand, we show the impact of vehicle automation with and without \ac{AV} lanes.
\end{itemize}

The remainder of this paper is organized as follows:
In \cref{sec:related} we discuss related work on \ac{AHS}.
\cref{sec:system} explains the system model used in this study.
In \cref{sec:analysis} we present our analytical evaluation of a synthetic scenario, followed by \cref{sec:case} in which results for the entire city of Singapore are presented and discussed.
\cref{sec:conclusion} concludes our article.

%

\section{Related Work}
\label{sec:related}

\ac{AHS} and their implications have received wide attention from both researchers and industry around the world.
Investigations include general AHS policies and concepts ~\cite{litman2016autonomous,kanaris1997spacing}, effects on travel times and capacity ~\cite{cohen2011impact,carbaugh1998safety,tsao1994capacity, harwood2014modelling,michael1998capacity,kanaris1997spacing, hall2005vehicle,van2006impact}, traffic safety~\cite{carbaugh1998safety,kanaris1997spacing, godbole2000safety, van2006impact}, and interactions between conventional human-driven vehicles and autonomous vehicles~\cite{dresner2007sharing}.

Several \ac{AHS} studies and field trials were conducted as part of the California \ac{PATH} program:
Tsao et al.~discuss the relationship between lane changing manoeuvres and the overall throughput of the highway~\cite{tsao1994capacity}.
Their analytical and simulation results indicate a direct trade-off between the two.
Lateral movement decreases the traffic flow, and higher traffic flow leads to longer lane change times.
In this work, we ignore decreased throughput caused by lane changing and focus on a best-case city-wide benefits of dedicated \ac{AV} lanes.
Godbole and Lygeros evaluate the increased capacity by the introduction of fully automated highways by treating the highway as a single-lane \ac{AHS} pipe~\cite{godbole2000safety}.
Similar to the work of Harwood and Reed~\cite{harwood2014modelling}, they study different platoon sizes and speeds but do not consider separated \ac{AV} and \ac{CV} lanes.

\ac{AHS} pipeline capacity was also studied by Michael et al.~\cite{michael1998capacity}.
Their results show that longer platoons of \acp{AV} are favourable as they increase the capacity of the road due to lower inter-platoon distances.
A high mixture of different vehicle classes, however, leads to a lower capacity.
Lastly, the presented analytical model shows that as highway speed increases, the capacity reaches a saturation point after which the speed decreases again.
This is inline with the findings presented in this paper when studying the throughput on only the \ac{AV} lane.

Similar to the work presented in this article, Cohen and Princeton investigated the impact of dedicated lanes on the road capacity~\cite{cohen2011impact}.
They find that assigning a dedicated lane can create bottlenecks due to increased lane changing manoeuvres of departing unauthorized vehicles.
When the capacity of the remaining lanes is exceeded, it may even be impossible for allowed vehicles to access the dedicated lane.
In our macroscopic study, we abstract away from these problems to answer the more general question whether the introduction of dedicated \ac{AV} lanes is a feasible approach.

In summary, it appears that while \ac{AHS} seem to be a well-studied subject, a general evaluation of the introduction of \ac{AV} lanes is still missing.
Furthermore, automated highways are mostly investigated in an isolated manner (and often also simplified to a 1-lane road) without taking into considering the rest of the road network.
Using the city-state  of Singapore as a case study, we show how travel times of both \acp{AV} and \acp{CV} are affected and that the introduction of dedicated AV lanes has a considerable effect on the entire road network by changing the distribution of traffic demand within the transportation system.

%

\section{System Model}
\label{sec:system}

The goal of this study is to evaluate the allocation of one lane on every highway road to be used exclusively by autonomous vehicles.
We investigate an increasing percentage of \acp{AV} in the system under two scenarios: with and without dedicated \ac{AV} lanes on highways.
In the former scenario, all roads that are not highways will exhibit normal traffic conditions as there will be a mixture of human drivers and AVs.
All lanes on the highway roads will be accessible to \acp{AV}, while \acp{CV} will be able to utilize all lanes except one lane, which will be allocated exclusively for \acp{AV} usage.

In the second scenario, all vehicles will share all lanes on all roads.

We model the different behaviour of \acp{AV} and \acp{CV} by means of smaller headway time, that is the time gap to the vehicle in front.
We assume that \acp{AV} can afford a much smaller headway than normal vehicles since their reaction time is orders of magnitude smaller than that of humans~\cite{ioannou1993autonomous}.
A direct consequence of this is that effectively the capacity of the road is increased, as \acp{AV} need less space.
The capacity (in cars per hour per lane) can then be calculated as follows:

\begin{equation}
C=\dfrac{3600}{h_{av}p_{av}+h_{cv}\left(1-p_{av}\right)}\label{capacity_equation}
\end{equation}

where $h_{av}$ is the headway time for AVs (values may vary between $0.5$ and $1$ second, depending on level of comfort~\cite{van2006impact}), $p_{av}$ is the percentage of AVs on the road segment, $h_{cv}$ is the headway time for conventional vehicles set to $1.8$ seconds.
Equation \ref{capacity_equation} is based on \cite{yokota1998evaluation}, where it was derived from collected data on highway roads in Japan for varying percentages of vehicles with \ac{AHS}.

\begin{figure}[h]
	\begin{center}

	{\includegraphics[width=0.6\columnwidth, trim=50mm 0mm 78mm 0mm, clip]{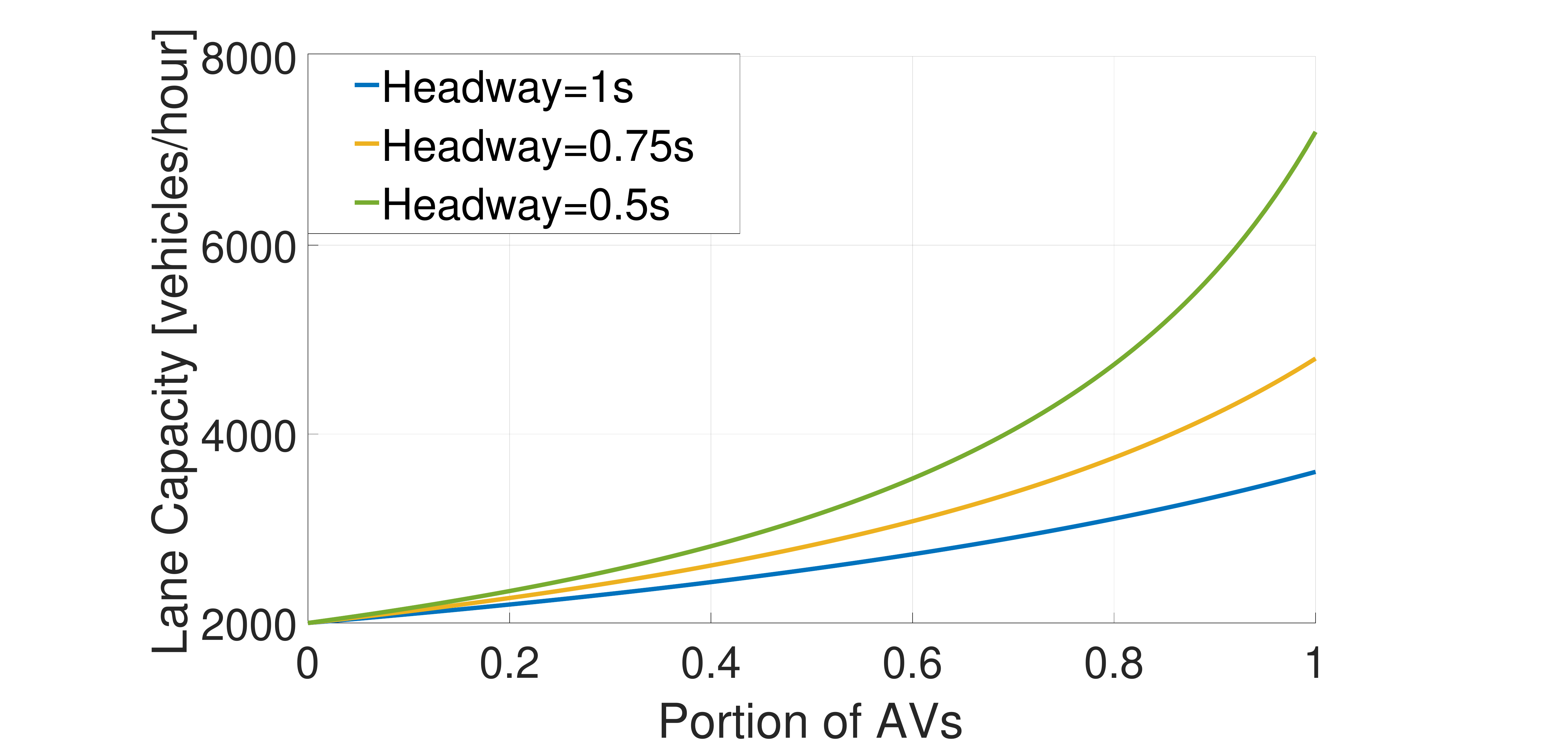}}
	\caption{Changes of road capacity as a function of AV percentage and headway time}
	\label{capacity}
\end{center}
\end{figure}

Figure \ref{capacity} illustrates the change of capacity as a function of the AV percentage for different AV headway values.
It can be observed that the capacity exponentially increases with the portion of AVs on the road.
This means that when there is a small portion of AVs on a road, their impact is only marginal, however, when the majority of vehicles are AVs there is a significant increase of the capacity of the road. The capacity for the same AV percentage also increases exponentially as the headway time decreases.
For the remainder of the paper we will assume a conservative \ac{AV} headway of $h_{av} = 1s$.

The primary measure we use to evaluate the impact on traffic caused by the introduction of an \ac{AV} lane is the travel time of cars.
The travel time $T = \sum_i^n t_i$ of a vehicle is determined by the traverse times of all $n$ segments (or links) included in its route.
The traverse time $t_i$ of a segment $i$ can be computed using the Bureau of Public Roads (BPR) function \cite{dafermos1969traffic}:\\
\begin{equation}
t_i=\dfrac{l_i}{{v}_i}\left(1+\alpha_i\left(\dfrac{F_i}{C_iw_it}\right)^{\beta_i}\right)\label{eq:bpr}
\end{equation}
where  $l_i$ is length of the road segment, ${v}_i$ is the free flow velocity of the segment, $F_i$ is flow, $w_i$ is the number of lanes, $t$ is time duration of the simulated period, $C_i$ is the capacity of road segment $i$, $\alpha_i$ and $\beta_i$ are parameters from the BPR function.
Free flow velocities $\hat{v}$ are extracted from historical GPS tracking data~\cite{sindhwani2010singapore}.
Parameters $\alpha_i$ and $\beta_i$ are calibrated for different classes of roads depending on their speed limits using both GPS tracking data and a travel time distribution of the population for certain periods of the day.
For a more detailed description of the calibration and validation procedures we refer the reader to \cite{t0zBWhqE4mBmv0Rm}.

%

\section{Analytical Evaluation}
\label{sec:analysis}

In this section, we will illustrate how to analytically determine the travel times of vehicles on highways in both cases with and without dedicated \ac{AV} lanes.
We examine a simplistic scenario, where the system consists of a single road with two lanes.
Let the overall percentage of AVs be $p$ and the percentage of AVs on the first and second lane respectively be $p_1$ and $p_2$. Let the number of vehicles on the respective lanes be $f_1$ and $f_2$ such that:
\begin{equation}
f_1+f_2=F \label{all_cars}
\end{equation}
where $F$ is the total number of vehicles in the system.
Therefore, we know that:

\begin{equation}
p_1f_1+p_2f_2=pF \label{probs}
\end{equation}
Under the assumption of \ac{UE}~\cite{fisk1980some} we also know that:
\begin{equation}
t(f_1)=t(f_2) \label{equality}
\end{equation}
where $t(f)$ represent the travel time on the road for flow $f$.
\ac{UE} is achieved when given the current traffic situation, a user would not change their route, i.e. the user is already on the shortest route in terms of travel time.

Substituting $t(f)$ for the BPR function (Equation \ref{eq:bpr}) we get:
\begin{equation}
t(f) = \dfrac{l}{v}\left(1+\alpha\left(\dfrac{f}{C}\right)^{\beta}\right)
\end{equation}
Where $\alpha$ and $\beta$ are parameters of the BPR function, $v$ and $l$ are the free flow velocity, and the length of the road and $C$ is the capacity of the road. 
Substituting the expression for capacity from Equation \ref{capacity_equation} we get for the BPR function:
\begin{equation}
t(f)=\dfrac{l}{v}\left(1+\alpha\left(\dfrac{f\left(h_{av}p_{av}+h_{cv}\left(1-p_{av}\right)\right)}{3600}\right)^{\beta}\right)
\end{equation}
Expanding Equation \ref{equality} we get:

\begin{eqnarray}
\dfrac{l}{v}\left(1+\alpha\left(\dfrac{f_1\left(h_{av}p_1+h_{cv}\left(1-p_1\right)\right)}{3600}\right)^{\beta}\right)=\nonumber\dfrac{l}{v}\left(1+\alpha\left(\dfrac{f_2\left(h_{av}p_2+h_{cv}\left(1-p_2\right)\right)}{3600}\right)^{\beta}\right) 
\end{eqnarray}

Removing common terms, simplifying and using Equations \ref{all_cars} and \ref{probs} we get:

\begin{eqnarray}
f_1\left(h_{av}p_1+h_{cv}\left(1-p_1\right)\right)=\nonumber\left(h_{av}\left(pF-p_1f_1\right)+h_{cv}\left(F-f_1-\left(pF-p_1f_1\right)\right)\right)
\end{eqnarray}

Let's examine the two scenarios in this simplistic example: In scenario $1$ a lane is exclusively reserved for AVs ($p_1=1$) and in scenario $2$ all cars are allowed to use all lanes and therefore get distributed equally on the two lanes ($f_1=f_2=\frac{F}{2}$ and $p_1=p_2=p$). For scenario $1$ we get:

\begin{equation}
f_1h_{av}=h_{av}\left(pF-f_1\right)+h_{cv}F\left(1-p\right)
\end{equation}

After simplifying we obtain:

\begin{equation}
f_1h_{av}=\dfrac{F}{2}\left(h_{av}p+h_{cv}\left(1-p\right)\right)\label{equal}
\end{equation}

Under the assumption of user equilibrium it follows that for scenario $1$ the travel time for both lanes is the same:

\begin{equation}
t_1=\dfrac{l}{v}\left(1+\alpha\left(\dfrac{f_1h_{av}}{3600}\right)^{\beta}\right)\label{scenario_1}
\end{equation}

The travel time for a single vehicle in scenario $2$ is:

\begin{equation}
t_2=\dfrac{l}{v}\left(1+\alpha\left(\dfrac{\dfrac{F}{2}\left(h_{av}p+h_{cv}\left(1-p\right)\right)}{3600}\right)^{\beta}\right)\label{scenario_2}
\end{equation}

It can be observed that because of Equation \ref{equal} the expression in Equation \ref{scenario_1} is equal to the expression in Equation \ref{scenario_2}.
This means that in the simple examined case, regardless whether a dedicated AV lane is introduced or not, the resulting travel time for all commuters is going to be the same after the point where there are enough AVs to ensure equilibrium can be reached.

The number of vehicles on the first lane in scenario $1$ is bounded from above by the total number of autonomous vehicles: $f_1 \leq pF$.
In the case where $f_1$ computed according to Equation \ref{equal} is smaller than $pF$ equilibrium cannot be achieved.
It is trivial to demonstrate that the overall travel time under scenario $1$ is bigger than the travel time under scenario $2$ when $f_1 \leq pF$.
The minimum number of AVs in terms of the AV percentage $p_{\tau}$ (the threshold percentage) in order for the solutions of the two scenarios to coincide is:

\begin{eqnarray}
p_{\tau}Fh_{av}&=&\dfrac{F}{2}\left(h_{av}p_{\tau}+h_{cv}\left(1-p_{\tau}\right)\right)\\
p_{\tau}&=&\dfrac{h_{cv}}{h_{av}+h_{cv}}\label{eq:tau}
\end{eqnarray}

The percentage $p_{\tau}$ describes the point where for an additional AV it would be equally attractive to use the AV lane or the normal lane. This is the moment where the AV lane reaches saturation.
Before that percentage is reached, the AVs have a shorter commuting time than the CVs, however, the system is performing worse in terms of overall congestion compared to the case without a dedicated AV lane.
Equation \ref{eq:tau} can be easily extended for an arbitrary number of lanes in our minimalistic example.
Assume we have $1$ dedicated AV lane and $N$ normal lanes.
Then the percentage of AVs after which the dedicated lane is saturated becomes:

\begin{equation}
p_{\tau}=\dfrac{h_{cv}}{Nh_{av}+h_{cv}}\label{threshold}
\end{equation}

Assuming headway times of $h_{cv}=1.8s$ and $h_{av}=1s$, the percentage $p_\tau$ of AVs on the road for saturation of the AV lane is $64.3\%$, $47.4\%$, $37.5\%$, $31\%$, $26.5\%$ for respectively $2$, $3$, $4$, $5$, $6$ lane roads with one allocated AV lane. 

%

\section{City-wide Simulation Study}
\label{sec:case}

We conduct a macroscopic city-wide simulation study of the city-state of Singapore to better understand the effects to be expected in a complex environment.
We take a look at the traffic conditions on the highway, but also closely investigate the effect dedicated \ac{AV} lanes have on the rest of the road network. 

\subsection{Methodology}

In order to evaluate the scenarios of AV introduction in a road transportation system, we make use of an agent-based macroscopic simulation approach.
The simulation consists of three steps: 1) Agent generation, 2) route computation and  3) travel time estimation.

The underlying road network on which the traffic assignment is performed is modelled by means of a uni-directional graph, where each edge represents a road segment and nodes represent decision points at which a road may split or merge.

To introduce dedicated \ac{AV} lanes, we alter the original graph by duplicating start and stop nodes of highway segments and creating a new edge to represent the \ac{AV} lane, while removing one of the normal lanes from the original segment.
Furthermore, costless connections are added between the start and end nodes and their respective copies so that vehicles can change lanes at the start/end of highway road segments.

Every agent is generated with an origin and destination sampled from an OD matrix representing the travel demand of the system.
Realistic traffic volume is modelled by synthesizing a sufficiently large vehicle population according to the available survey data. 

The routes of the agents are computed using an incremental user equilibrium approach~\cite{fisk1980some} aiming at representing reality in the sense that every driver is satisfied with their route and would not choose a different one given the current traffic situation.
We assume a driver is satisfied when they are on the shortest route from origin to destination in terms of travel time.
Routing is performed on the road network graph, where each edge has an attached weight representing the current traverse time of this road segment.
Weights are updated after every batch route computation.
To disallow conventional vehicles from using the \ac{AV} lane, the weight of the edges representing these lanes are set to infinity when the route of a \ac{CV} is computed and set back to their traverse time values for \ac{AV} route computation.

Further assumptions regarding to the macroscopic level of the simulation include:
\begin{itemize}
	\item Agents want to minimize travel time and all perceive the traffic situation in the same way (there is no noise in the observations of the traffic states).
	\item Traffic is spread homogeneously in time during the simulation.
	In order for the BPR function (Equation \ref{eq:bpr}) to give a reasonable estimation of the traffic conditions, the flow $F$ during the simulation time $t$ needs to be spread homogeneously in time.
	If many vehicles try to enter the road at the same time for example, the traffic congestion generated will likely produce a much higher traverse time for the road than the expected one.
	In order to ensure as much as possible that traffic demand stays static throughout the simulation period, both the period $t$ and the time of day, which is simulated, have to be chosen with care.
	\item Agents do not reroute while on their trip.
	In reality drivers may change their trip plan dynamically according to unexpected events or observed traffic conditions ahead.
	As our traffic assignment provides information to the drivers about expected traffic conditions for the whole network, we assume that such events will not occur as drivers are given all the information they need prior to their trip.
\end{itemize}

\subsection{Data and Scenario Description}

We examine the city-state of Singapore with a total population of $5.4$ million and around $1$ million registered vehicles including taxis, delivery vans and public transportation vehicles.
The fact that Singapore is situated on an island simplifies our scenario as the examined system is relatively closed with only two expressways leading out of the country.
There are $3495$ km or roads of which $652$ km are major arterial roads and $161$ km are expressways (these are the roads on which we introduce the dedicated AV lane indicated on \cref{fig:highways}) spreading over $715$ squared km of land area.
We have used publicly available data to acquire a unidirectional graph of Singapore, that comprises of $240,000$ links and $160,000$ nodes representing the road network of the city.
The number of lanes, speed limit and length of every link is available allowing us to extract information about its capacity.

For the purposes of our model we make use of two separate data sets.
The first data set consists of GPS trajectories of a $20,000$ vehicle fleet for the duration of one month, providing information about recorded velocities on the road network during different times of the day~\cite{sindhwani2010singapore}.
The second one is the Household Interview Travel Survey (HITS) conducted in $2012$ in the city of Singapore, which studies the traffic habits of the population.
Information about the origin destination pairs, their temporal nature, and commuting time distribution during rush hour periods is extracted from it.

In order to achieve realistic traffic conditions, we estimated the number of agents based on the 
Singapore HITS data.
We extracted the ratio of people who actively create traffic on the streets (cab drivers, personal vehicle drivers, lorry drivers) and the total number of people interviewed.

We chose the morning commute hours (7:30am - 8:30am) as the period which seems to be the most stable in terms of traffic volume and estimated the traffic demand consisting of $309,000$ agents.

Similar to the simplistic example the vehicle population is split into two parts: \ac{AV} and \ac{CV}.
The percentage of \ac{AV} is varied in order to observe the effects in the initial stages of \ac{AV} introduction, as well as the possible traffic situation if all vehicles were autonomous.

In order to have a benchmark for measuring the efficiency of the suggested policy, we also evaluate the scenario where no \ac{AV} lane is introduced and all vehicles can access all lanes on all roads.
The analytical solution acquired for the simplistic example in the previous section indicates that the scenario with no \ac{AV} lane will produce better or at least the same system performance as the introduction of the \ac{AV} lane. 
Our case study will test our analytical results for a more realistic transportation system environment.

\subsection{Effects on Average Travel Time}

We evaluated the change in average commute time based on the percentage of \acp{AV} in the system.
We compare two different settings: with and without a dedicated \ac{AV} lane.
As a baseline we use the average commute time without the existence of \acp{AV} (and thus no exclusive lanes).
This time was found to be approx. 18.5 minutes and can be seen as the status quo.

We simulated both settings with the exact same travel demand, that is, identical origin-destination pairs for all vehicles.
We then gradually increased the percentage of \acp{AV} in the system.
The number of vehicles in the simulation was invariant; a higher percentage of \acp{AV} means that \acp{CV} were replaced with autonomous vehicles.

\begin{figure*}[h]         
	\centering
	
	\subfloat[Travel times with dedicated AV lane]{
		\includegraphics[trim={150 0 150 0}, clip,width=.45\textwidth]{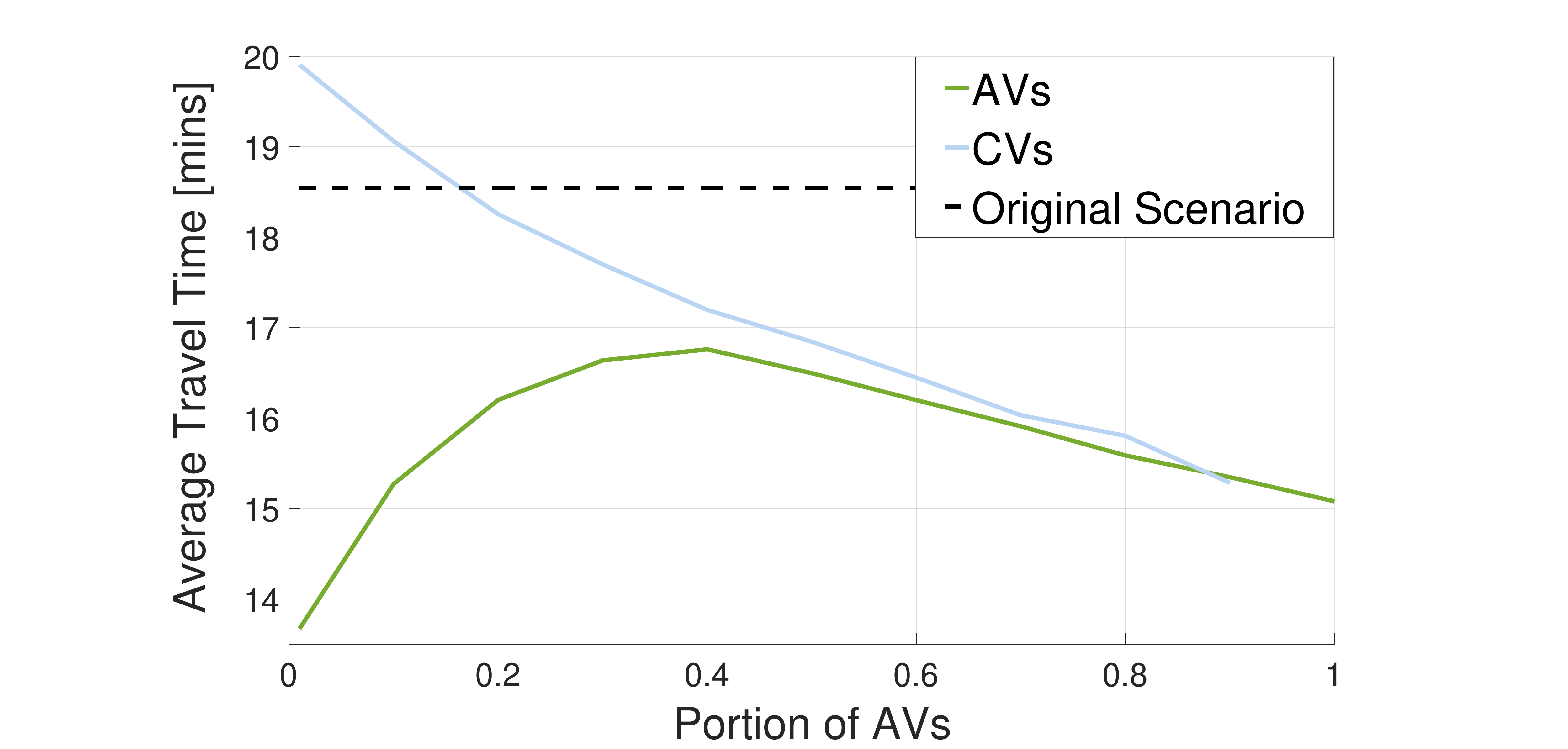}\label{fig:with_lane}
	}
	\qquad
	\subfloat[Comparison for all vehicles]{
		\includegraphics[trim={150 0 150 0}, clip,width=.45\textwidth]{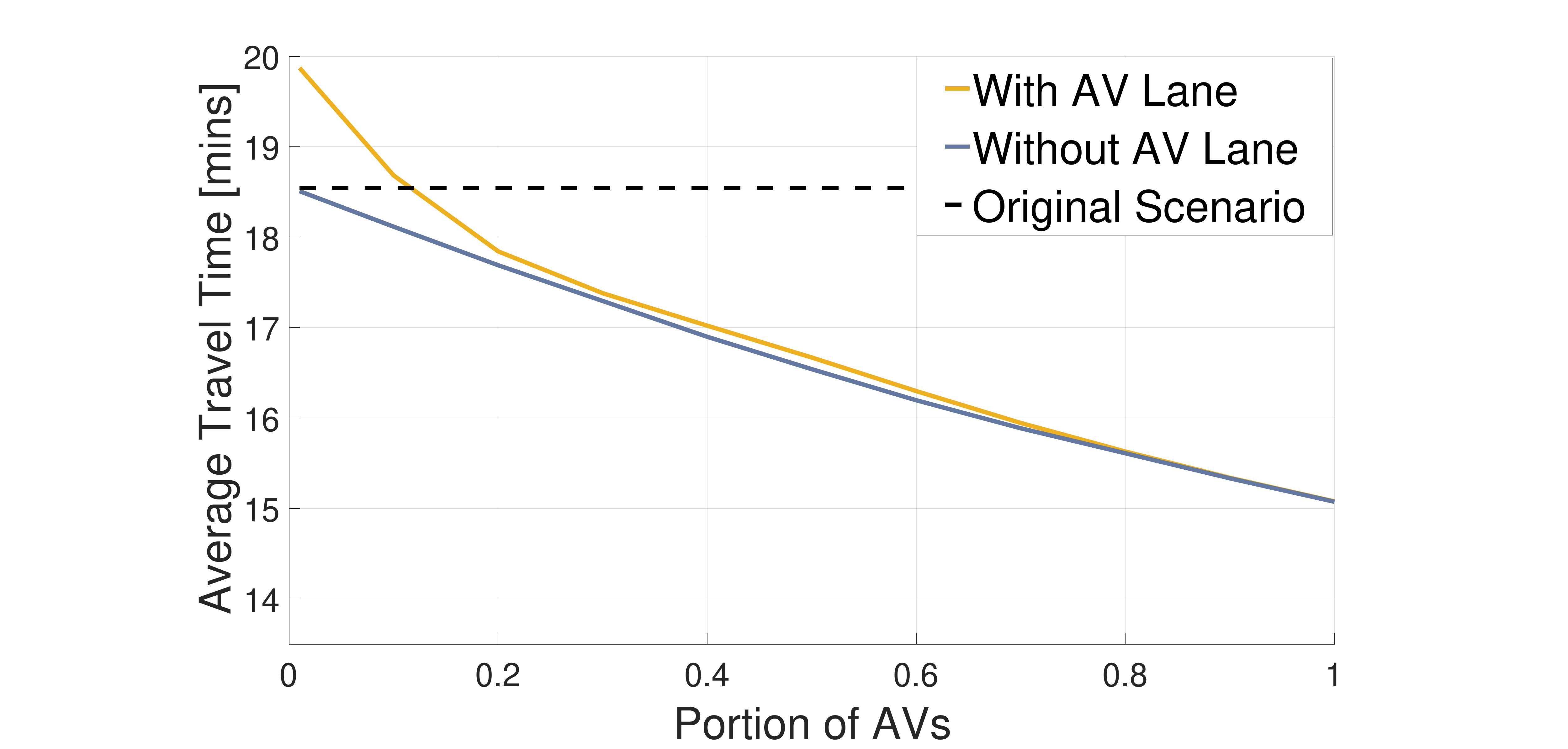}\label{fig:comparison}
		
	}
	
	\caption{Average travel time of whole population, AVs and CVs as a function of AV percentage}
	\label{fig:travel_time}
\end{figure*}

\cref{fig:travel_time} shows our result.

The black dashed line serves as an illustration of the status quo.
It can be observed that when introducing a lane exclusively for \acp{AV} and thereby taking away capacity from conventional vehicles (\cref{fig:with_lane}), the travel time for \acp{CV} increases, whereas the low percentage of \acp{AV} (driving on almost empty lane) travel considerably faster.
This could be used as an incentive by policy-makers to increase the share of \acp{AV} in the transportation system.
With an increasing percentage of \acp{AV} we observe that travel times for \acp{CV} decrease due to the fact that effectively vehicles move away from the common lane to the AV lane, which in turn increases the travel time on the AV lane.
This behaviour can be observed until the AV lane is saturated (somewhere between 40\% and 50\% AV percentage).
From this point on, the choice of lane makes no difference for newly added \acp{AV} as they experience equal travel times regardless whether they choose to take the \ac{AV} lane or the mixed lane. 
Therefore, the difference between average travel time of \ac{AV} and \ac{CV} decreases. 
Travel time of both \acp{AV} and \acp{CV} still decreases with the increase of the percentage of \acp{AV} since the capacity of the road network is effectively increased.

The average number of lanes in the Singapore road network is $3.14$.
If we apply Equation \ref{threshold} from our analytical evaluation, we yield a saturation point of $45.2\%$ AV penetration, which seems to be in agreement with our simulation results.

We conclude that computing the threshold percentage is useful in terms of AV lane introduction planning.

Figure \ref{fig:comparison} shows the comparison between the average travel times for the entire vehicle population with and without the introduction of an AV lane. 
It can be observed that the setting without an AV lane is always performing better than the AV lane one. 
After saturation of the \ac{AV} lane the difference between the two curves becomes marginal.
This finding is in complete agreement with the results of our analysis in \cref{sec:analysis}. 
Before saturation of the \ac{AV} lane, the capacity of the highway is not fully utilized and that, on average, the smaller travel times of the \acp{AV} cannot make up for the introduced delays for the \acp{CV}.
We conclude that adding an \ac{AV} lane, while initially being an incentive for early adopters, will noticeably penalize drivers of conventional vehicles at the early stages of \ac{AV} introduction.
Additional delays will be introduced by lane changing manoeuvres and other microscopic effects not considered in this macroscopic study~\cite{tsao1994capacity,cohen2011impact}.

\subsection{Analysis of Effect of Headway Time}

The headway time of vehicles is a crucial parameter for the computation of the road capacity (see Equation \ref{capacity_equation}) and therefore an important input for the travel time computation using the BPR function (see Equation \ref{eq:bpr}).

Although, AVs can afford to have smaller headways, this might negatively affect the comfort of the passengers, as small distances at high velocities can induce anxiety~\cite{van2006impact}. 
To this end, we examine the effect of varying the headway time (from $0.5$s to $1$s) for the city-scale simulation scenario with a dedicated AV lane on highways. 
It is therefore useful to evaluate the amount of overall travel time decrease as a function of the headway time. 

\begin{figure}[h]
	\begin{center}
	{\includegraphics[width=0.6\columnwidth, trim=150 0 150 0, clip]{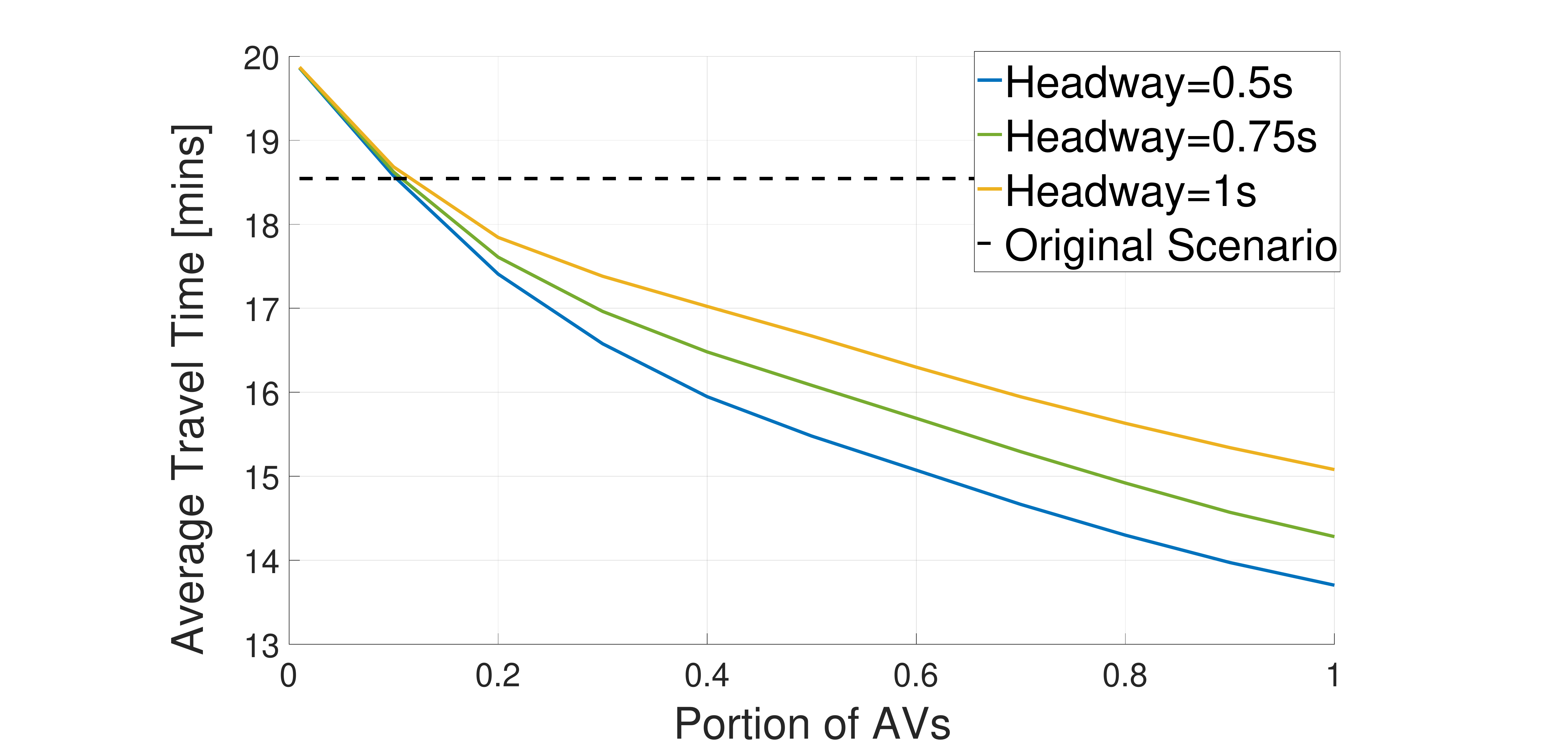}}
	\caption{Average travel time curves for varying headway time}
	\label{fig:headway}
	\end{center}
\end{figure}

In \cref{fig:headway} we observed that, depending on the headway, the improvement of overall travel time can vary between $20\%$ and $26\%$. 
The difference between the improvement in the case of all AVs seems is bigger between headway $0.5$s and $0.75$s than between $1$s and $0.75$s.

This is due to the non-linear nature of the BPR function.
As vehicles approach free flow velocities due to the increased capacity, any further improvement in capacity plays a smaller role and therefore the gains become less significant. 
It must be noted, that this happens only when traffic congestion is such that vehicles travel at free flow velocities.

If the system was more congested, we would not be able to observe the decrease of travel time improvement for smaller headway time, since the BPR function will still be in its non-linear part thus providing significant changes of travel time as the capacity is altered.

\subsection{Effects on Fuel Consumption}

We have demonstrated that the introduction of AVs decreases the overall travel time in the system. 
The next question we are trying to answer is by how much this saved time reduces fuel consumption (or energy in the case of electric vehicles). 
We deploy the Elemental fuel consumption model~\cite{faris2011vehicle} to evaluate the average fuel consumption of the vehicle population and subgroups.

\begin{figure}[h]
	
	\begin{center}
		{\includegraphics[width=0.6\columnwidth, trim=145 0 150 0, clip]{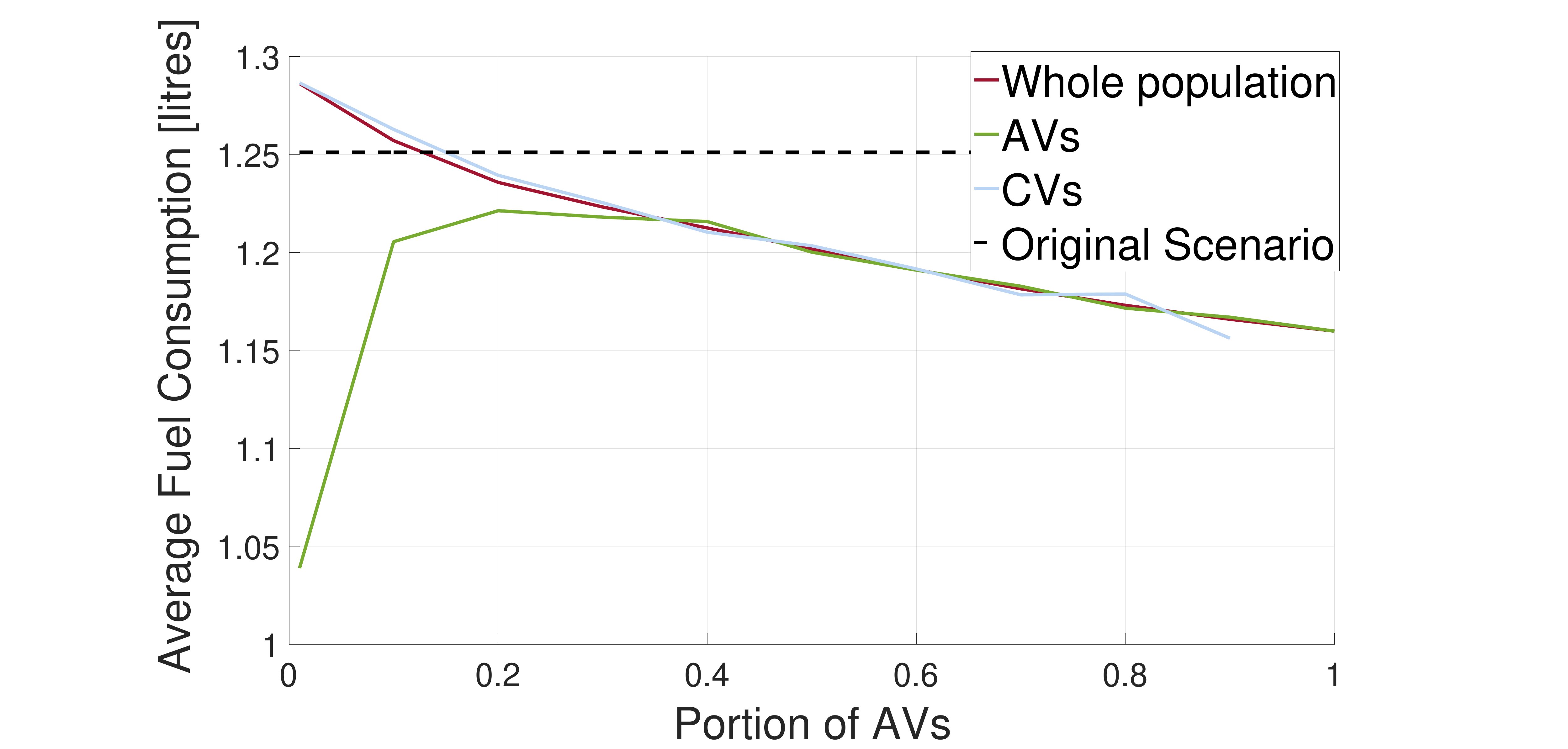}}
		\caption{Average fuel consumption of whole population, AVs and regular vehicles as a function of AV percentage}
		\label{fig:fuel_consumption}
	\end{center}
	
\end{figure}

The effects of introducing an AV lane on fuel consumption are shown in \cref{fig:fuel_consumption}

As expected, we observe a positive effect on fuel consumption.
To provide an idea of magnitude, the fuel saved at 20\% AVs is about $9.3$ tons for the duration of one Singapore morning rush hour and approx.\ $45$ tons when every vehicle is autonomous.
As for the travel time, the combination of one dedicated lane and a low percentage of \acp{AV} penalizes \acp{CV}.

The introduction of more AVs, however, decreases both the overall travel time and fuel consumption. 
It can also be noted that the relative effect of the fuel consumption is smaller than the effect on travel time, which means that fuel consumption is less sensitive with respect to changes in the traffic states for the examined scenario.
Please note that we do not consider fuel or energy saving resulting from platoon organization~\cite{hall2005vehicle} and other AV related technology but only seek to provide a order of magnitude with our macroscopic evaluation.

\subsection{Effects on Road Network Throughput and Traffic Distribution}

\begin{figure}[!t]         
	\centering
	\subfloat[Singapore road network with highways featuring dedicated AV lanes in bold]{
		\includegraphics[trim={150 0 150 0}, clip,width=0.5\columnwidth]{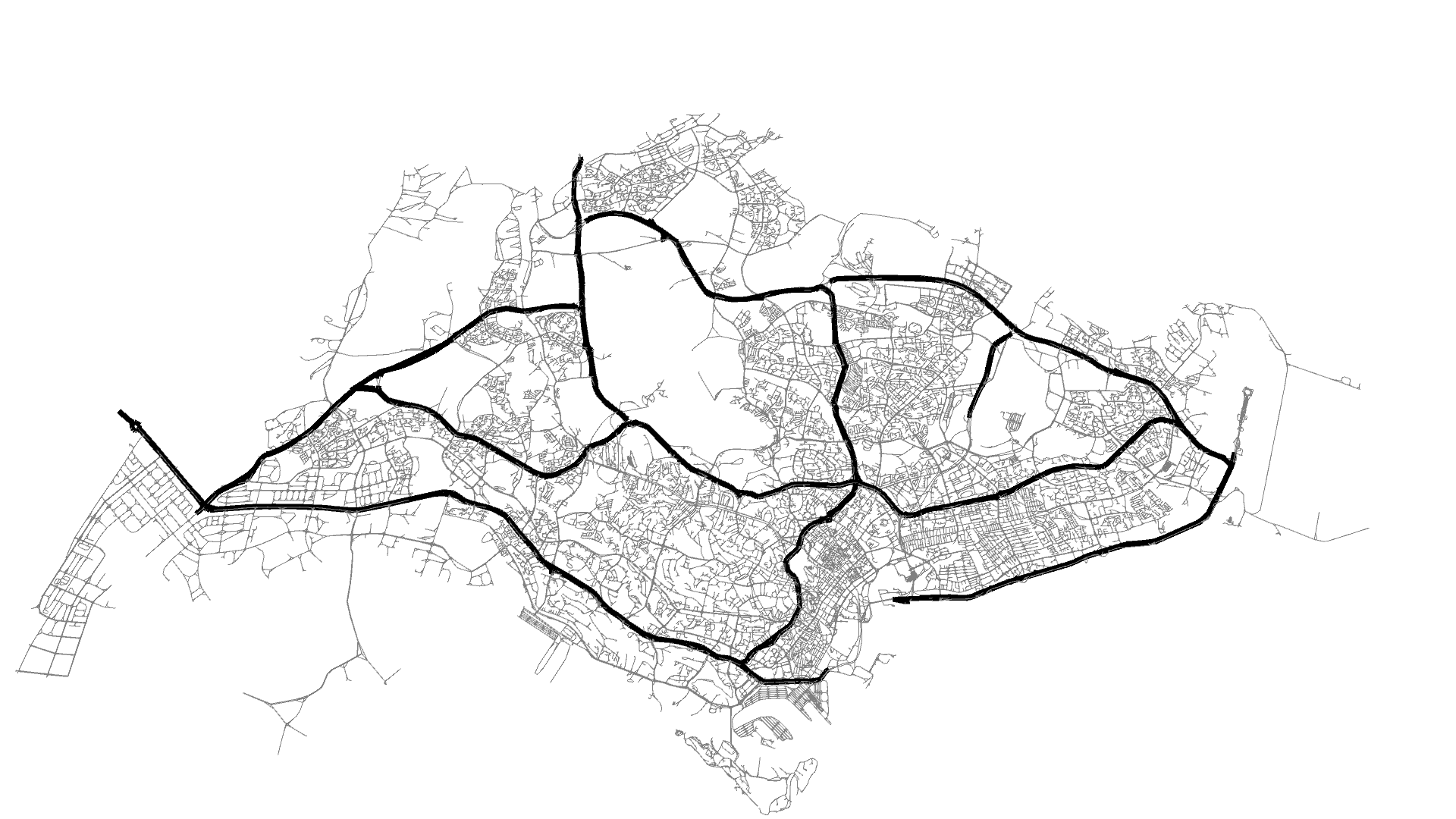}\label{fig:highways}
	}\qquad    	
	\subfloat[Change of road throughput at 50\% AVs (with AV lane)]{
		\includegraphics[trim={150 0 150 0}, clip,width=0.5\columnwidth]{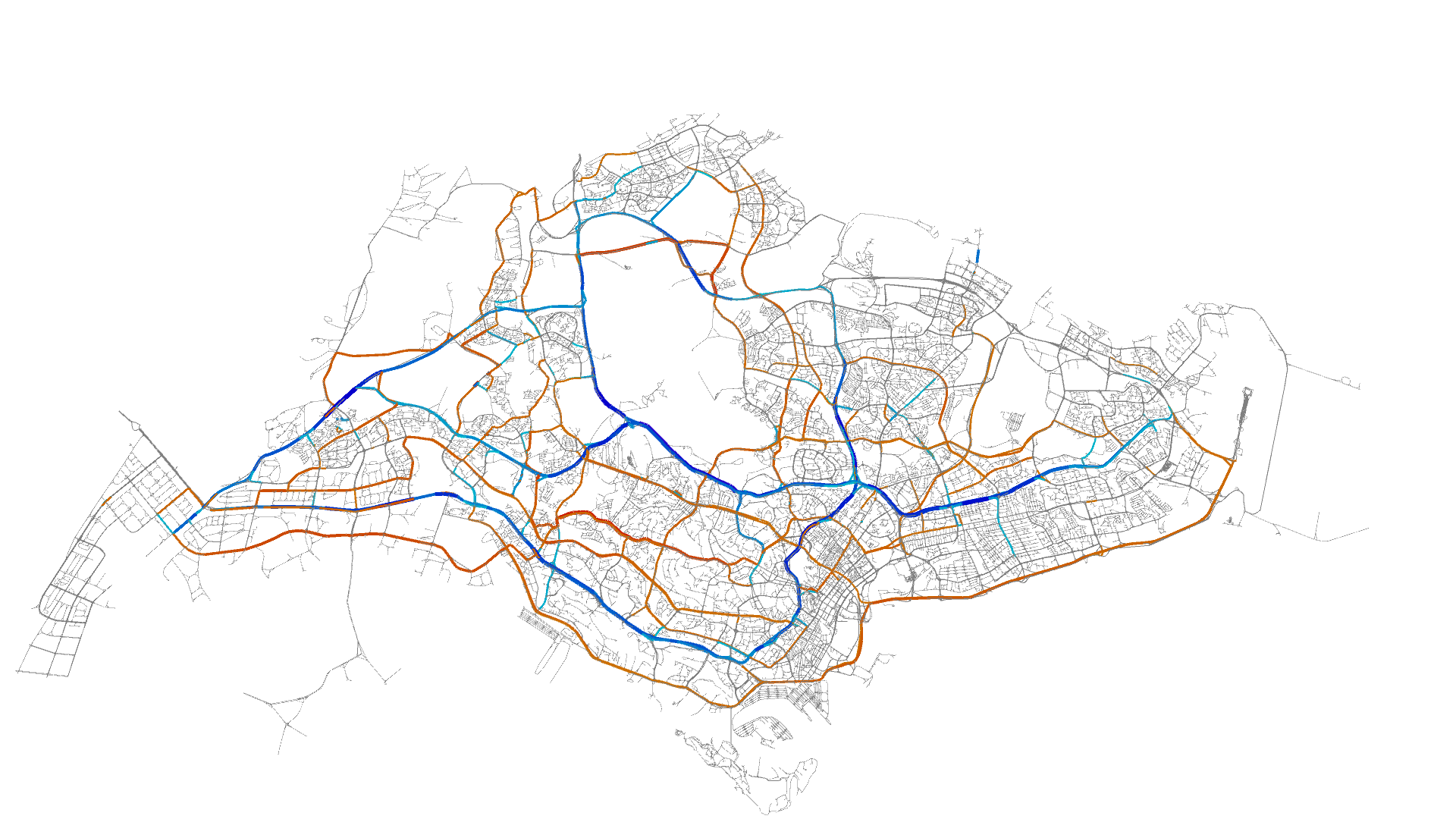}\label{fig:with_through}
	}\qquad
	\subfloat[Change of road throughput at 50\% AVs (without AV lane)]{
		\includegraphics[trim={150 0 150 0}, clip,width=0.5\columnwidth]{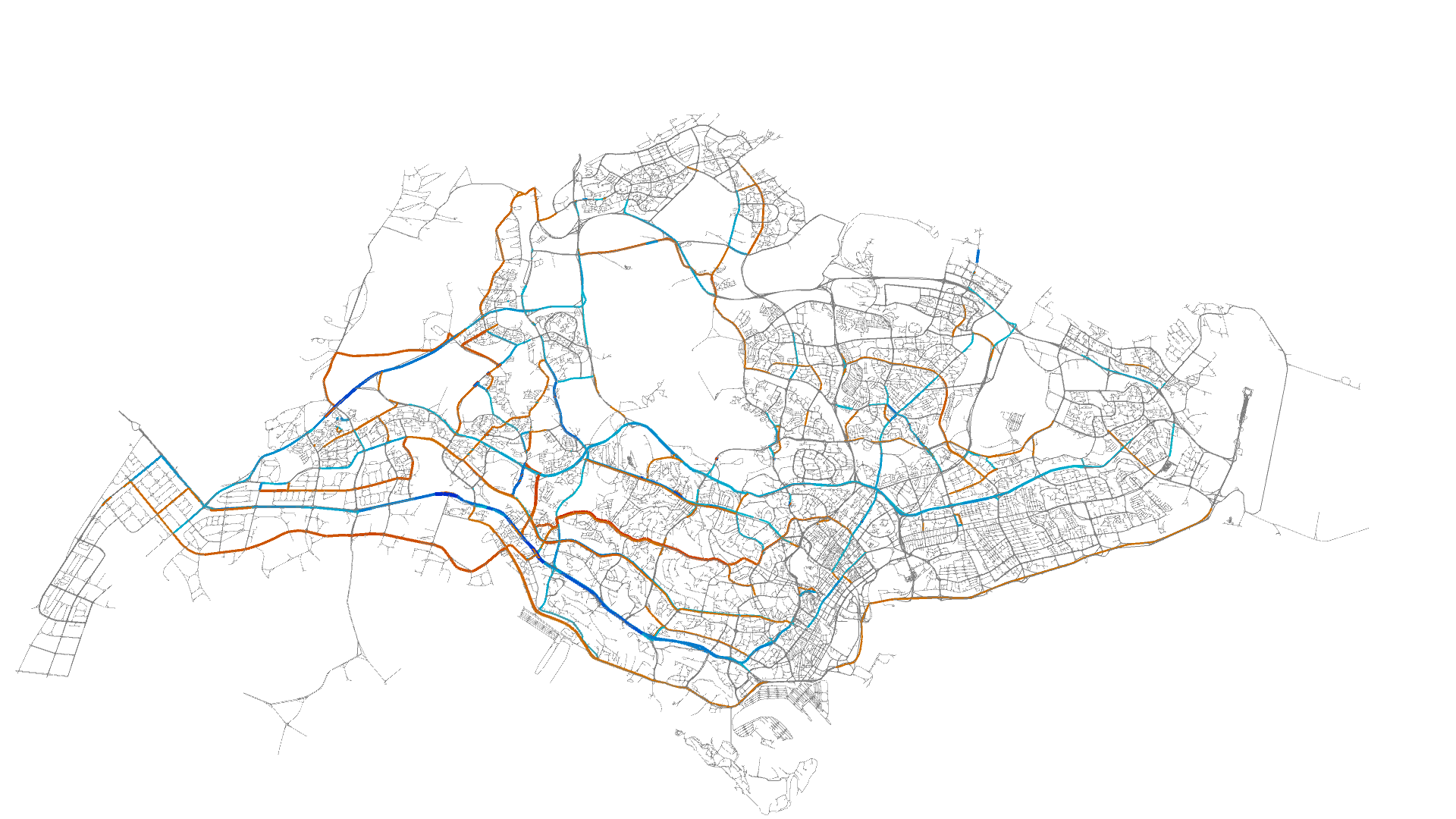}\label{fig:no_through}
	} 	
	\caption{Road throughput change caused by 50\% AVs compared to 0\% AVs. Colours from the blue gamma represent higher throughput; colours in the red gamma represent lower throughput}
		\label{throuput_change}
\end{figure}

\begin{figure}[h]
	\centering
		\includegraphics[trim={120 0 120 0}, clip,width=0.6\columnwidth]{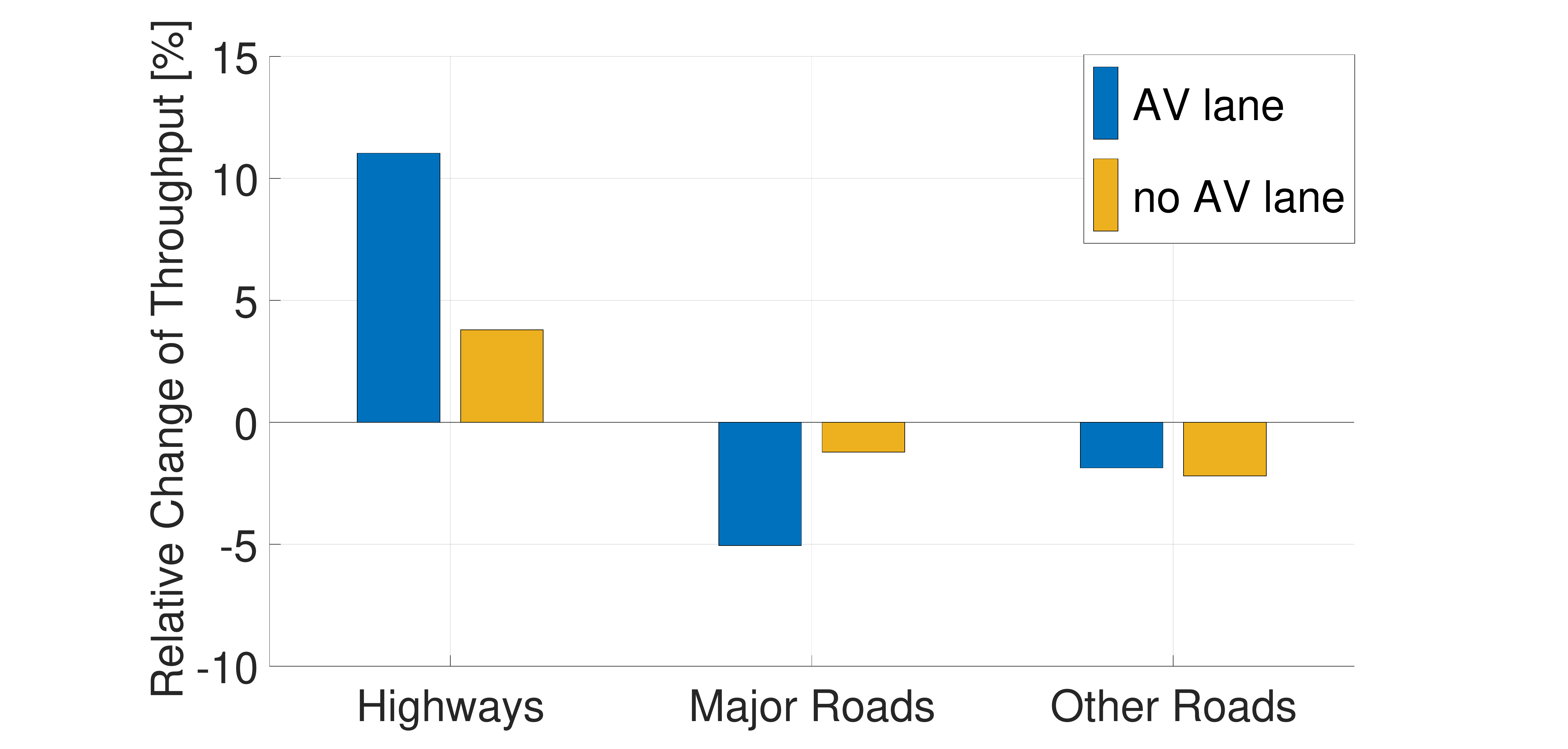}
\caption{Change of road throughput for different road types}
\label{fig:hist}
\end{figure}

Lastly, we examine the traffic distribution on the entire road network to gain a better understanding of the changes occurring in traffic conditions as a results of the introduction of dedicated AV lanes.
To the best our knowledge, existing literature focuses on the traffic changes on the highways and ramps only, and such a study has not been conducted before.
\cref{fig:highways} shows the Singaporean road network with highways drawn in bold.
AV lanes are added only to these highways, the rest of the network remained unchanged.

As a first step, we provide a qualitative measurement by visualizing the impact an introduction of $50\%$ \acp{AV} has on the road network (compared to $0\%$ AVs).
To this end, we draw roads experiencing higher throughput in blue colours and roads experiencing lower throughput in red colours.
We show our results for both settings, i.e., with AV lane (\cref{fig:with_through}) and without AV lane (\cref{fig:no_through}).

It can be observed that in both cases the highways exhibit a considerably higher throughput of vehicles, which is expected since the capacity of the highway is technically increased by the introduction of the autonomous vehicles.
Other roads exhibit a slight decrease of throughput, which may be due to the changes in routing that are triggered by the AVs, which have a strong preference towards the highways. 
In other words, the AVs are, in a way, attracted to the highways, since they can traverse fast there and the capacity is sufficient. 
This, however, makes them willing to pass through more congested roads in order to get to the highway, which creates additional time losses for the regular vehicles. 
This argument is further strengthened by the fact that the roads leading to the highways do not exhibit increased throughput, although they exhibit higher demand since more vehicles want to use the highways. 
This means that the level of congestion on those roads is too high to allow an increase of throughput.

Comparing the throughput changes in the two scenarios, it can be observed that the increase of throughput on the highways for the case with no AV lane is smaller. 
Therefore, the change of routing triggered by the introduction of AVs is qualitatively the same but quantitatively different for the two examined scenarios.
This can be observed in Fig. \ref{fig:hist}.
The relative increase of throughput for highways is almost three times higher for the dedicated AV lane case.
Higher level of throughput increase on highways leads to higher level of throughput decrease on the major roads, which represent alternative routes.
As mentioned before, if highways become too attractive for AVs, there can be negative effects on traffic conditions stemming from overly utilized roads which lead to the highways. 
The more balanced distribution of traffic in the no AV lane scenario could be the reason for its slight, however, consistent superiority over the dedicated AV lane case in terms of average travel time observed in \cref{fig:comparison}.

Finally, we investigate the change in demand for different types of roads.
We define demand for a road as the number of vehicles that have this particular road in their route under user equilibrium traffic assignment.
Taking a closer look at the difference in travel demand between the scenario with AV lane and the scenario without the introduction of AV lane, we measure the relative difference of travel demand between the scenarios.
Formally, let the demand for road $i$ for a given AV percentage $k$ be $D_{i,k}^{a}$ for the AV lane scenario and $D_{i,k}^{b}$ for the benchmark scenario without an AV lane.
Then, for the different classes of road (highways, major roads, other roads), we compute the difference in demand between the two examined scenarios relative to the AV lane scenario (a negative value therefore means that this particular type of road experiences less demand when AV lanes are introduced compared to not introducing AV lanes): 

\begin{equation}
\Delta_{D,k}=\dfrac{\sum\limits_{i}D_{i,k}^a-\sum\limits_{i}D_{i,k}^b}{D_{i,k}^a}
\end{equation}

\begin{figure}[h]
	\centering
	\includegraphics[trim={120 0 150 0}, clip,width=0.6\columnwidth]{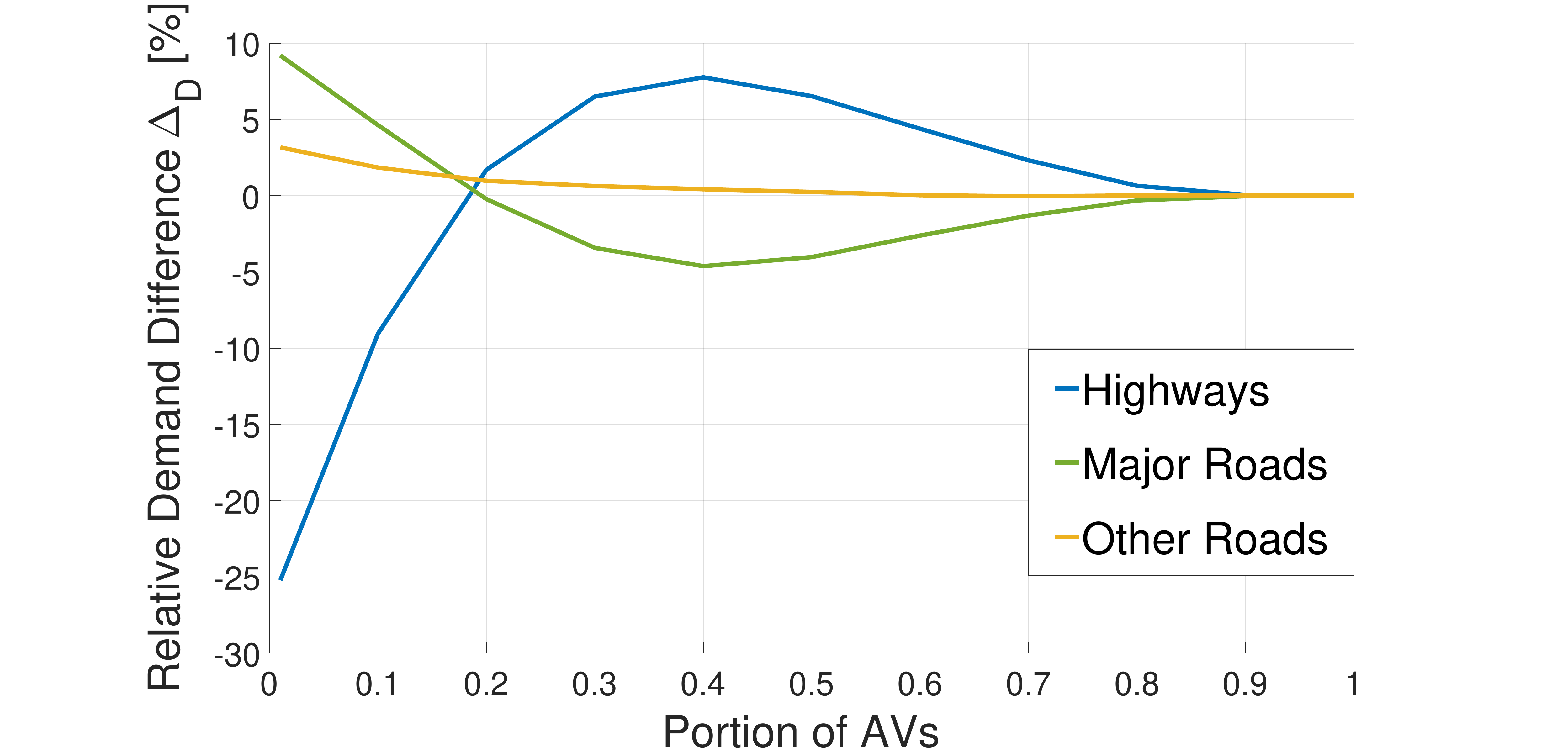}
\caption{Demand difference between AV lane setting and no AV lane setting for varying percentage of AVs}
\label{fig:demand}
\end{figure}

\cref{fig:demand} shows our results for highways, major roads, and other roads. 
For highways, we observe that initially there is lower demand.
This is due to the lower road capacity resulting from the dedicated lane, causing more \acp{CV} to avoid the highways.
With more \acp{AV} in the system, the demand for the highways increases until the saturation point of the AV lane is reached.
From this point on, the difference between the both settings becomes smaller, eventually converging to zero as the addition of more \acp{AV} cause the common lanes to achieve capacity values close to the one of the AV lane.

For the major roads we observe increased demand as \acp{CV} will favour these roads as an alternative to the more congested highways caused by the dedicated AV lane.
This effect decreases until the saturation point of the AV lane.

At this moment the highways reach their maximum demand difference and therefore a smaller portion of the population takes the alternative routes utilizing major roads.
Following the negative peak, the demand difference for major roads also converges to zero as the two scenarios become identical.

The difference for smaller roads is less pronounced and can only be observed at a low percentage of \acp{AV} in the system.
The higher utilization of major roads will also increase traverse time of these roads.
Some vehicles will therefore choose to use minor roads as a third alternative.

It is interesting to note that after the saturation point of the AV lane the difference in travel times between the two scenarios is almost negligible, however, the actual assignment of traffic, as can be observed in \cref{fig:demand}, is qualitatively different.
This finding indicates that the introduction of an AV lane would not just affect the travel time of the population but also shift the route preferences of commuters.

\section{Conclusion and Future Work}
\label{sec:conclusion}

In this article we demonstrated the effect of assigning one lane on highways exclusively for \acp{AV}.
We showed that for lower percentages of \acp{AV}, or more precisely, before the dedicated lane is saturated, travel times for \acp{AV} can be significantly shorter, while at the same time \acp{CV} are delayed due to the reduced capacity of the highway.

Looking at the entire road network, we observe that also non-highways are affected as \acp{CV} will effectively be drawn away from the highways onto the major roads.
This effect is especially pronounced at early stages of AV adaptation where the AV lane will remain mostly empty.
Regardless of an introduction of the \ac{AV} lane, we confirmed earlier findings that a larger number of \acp{AV} will have a positive impact on travel times and fuel consumption for all vehicles.
The macroscopic simulation study confirmed our analytical evaluation where we showed how to compute the saturation point of the \ac{AV} lane and illustrated that after this point is reached, the benefits for \ac{AV} users become negligible.
We further compared the scenario with \ac{AV} lane introduction to a baseline scenario where no changes to traffic regulations are made.
The latter scenario outperforms the former one over the whole range of \ac{AV} percentages, however, the difference is of considerable amount only before the saturation point is reached.
This finding coincides with our analytical evaluation of the two scenarios.

Future work includes micro (and submicroscopic) studies to further provide insights on the effects of the introduction of AV lanes.
This includes benefits due to smart platooning strategies but also turbulences caused by lateral vehicle movement, e.g., from the on-ramp towards the AV lane.
Furthermore, the authors would like to exchange the \ac{UE} traffic assignment of the \ac{AV} group of vehicles with the BISOS algorithm ~\cite{7795902}, which looks for a system optimum assignment, in order to check whether the adoption of an \ac{AV} lane would become more beneficial in this case.

\section*{Acknowledgements}
This work was financially supported by the Singapore National Research Foundation under its Campus for Research Excellence And Technological Enterprise (CREATE) programme.

\vspace{-1em}
\bibliographystyle{IEEEtran}
\bibliography{bibliography}

\end{document}